\documentclass[aps,prl,reprint,amsmath,amssymb,longbibliography,nofootinbib,superscriptaddress,floatfix]{revtex4-2}
\usepackage[T1]{fontenc}
\usepackage{lmodern}
\usepackage{microtype}
\usepackage{mathtools}
\usepackage{booktabs,array}
\usepackage{tikz}
\usepackage{hyperref}
\hypersetup{colorlinks=true,linkcolor=blue,citecolor=blue,urlcolor=blue,pdfauthor={Laurentiu Rodina},pdftitle={Beyond hidden zeros and uniqueness in Tr(phi^3): Block and rod zeros in any Tr(phi^m)}}
\newcommand{\tr}{\operatorname{Tr}}
\newcommand{\rank}{\operatorname{rank}}
\newcommand{\G}{\mathcal G}

\newcommand{\calD}{\mathcal D}

\newcommand{\ee}{\varepsilon}

\usepackage{xcolor}

\begin{document}
\title{Beyond hidden zeros and uniqueness in $\tr(\phi^3)$: Block and rod zeros in any $\tr(\phi^m)$}
\author{Laurentiu Rodina}
\email{laurentiu.rodina@gmail.com}
\affiliation{Beijing Institute of Mathematical Sciences and Applications, Beijing, China}
\date{November 19, 2026}

\begin{abstract}
Hidden zeros have proven powerful tools for reconstructing amplitudes in $\tr(\phi^3)$, a theory of colored scalars with cubic interactions. We show that these structures extend to any $\tr(\phi^m)$ theory. For odd $m$, we construct \emph{block zeros} from orthogonal sums of consecutive momenta, generalizing the original cubic hidden zeros. For every $m\geq4$, including even $m$, we describe a second family of nonlinear \emph{rod zeros} through a rod packing problem. In $\tr(\phi^4)$ this construction reduces to a simple rank-one condition on a matrix of momentum contractions. We connect this new zero to the classic threshold nullification zeros of Voloshin, where certain amplitudes in massive $\phi^4$ theory vanish when all produced particles are at rest. Finally, we demonstrate the reconstruction power of these zeros. Assuming locality, we prove that the rod zeros uniquely fix all $\tr(\phi^m)$ tree amplitudes for $m\geq4$ up to normalization. We also conjecture that the block zeros enforce both locality and unitarity for any $\tr(\phi^m)$ theory with odd $m$.
\end{abstract}
\maketitle

\section{Introduction}
\label{sec:intro-main}
Hidden zeros are kinematic loci on which scattering amplitudes vanish while every propagator remains finite. In $\tr(\phi^3)$, the nonlinear sigma model (NLSM), and Yang--Mills theory~\cite{hidden}, they arise from nontrivial cancellations between diagrams that are not apparent from the usual Feynman expansion. Requiring an amplitude to vanish on such a locus relates terms with different poles, and can therefore constrain both the allowed pole combinations and their residues.

This gives hidden zeros a natural role in the uniqueness program, which asks how much of an amplitude can be determined from a small set of principles. In a conventional tree construction, locality determines which poles may occur together, while unitarity fixes their residues through products of lower-point amplitudes. Reconstruction from gauge invariance, soft behavior, and scaling at large complex momenta has shown that these properties need not always be assumed independently~\cite{gaugeunitarity,gaugeunique,bcfwunique,bosesymmetry,softbootstrap,softuniqueness,uvweb,chyunique}.

Hidden zeros realize this possibility in tree-level $\tr(\phi^3)$, where they enforce locality~\cite{locality} and unitarity~\cite{cubicunique} at every multiplicity within a prescribed planar pole ansatz. Their constraining power extends to loop integrands, where, assuming locality, tree-boundary zeros together with zeros from Kontsevich's matrix Airy function~\cite{kontsevich} determine relative factorization and cut data in the loop cases studied~\cite{oneloop,multiloop}. Reconstruction from zeros has also been developed for pion amplitudes, cosmological wavefunctions or inverse Kawai--Lewellen--Tye (KLT) amplitudes~\cite{nlsmzerorecursion,cosmounique,classical}, with further examples of hidden zeros found in particle and string theories and form factors~\cite{expansionzeros,stringyklt,largedeformations,massivezeros,fermioniczeros,stringformfactors,formfactorsplit,allstringzeros,integerrectangles,openstringrepresentations,cegmsplit,cegmnlsm,higherderivativezeros,cosmozeros,inverseklt,kltintersection,stringyklt,inversekltgeometry,inversekltsurface}. The relation to lower-point functions becomes explicit in splitting relations motivated by zeros~\cite{smoothsplit,universalsplit,universalsplitdetail,allsplits,ymfactorizations,ymfactorizationproof,onshellsplitting,bcfwsplit,diagrammaticsplitting,pionexact,diagramtwo,universaldiagrams,onshellrepresentation,higherderivativesplit,loopfeynmanzeros,splitting}. These cancellations also have a classical explanation based on the equation of motion of the scalar theory~\cite{classical}.

Despite this range of examples, many of the structures found in other theories remain connected to $\tr(\phi^3)$ through kinematic shifts, scalar scaffolding, and double copy~\cite{nlsmembedding,circlestriangles,scaffoldedgluons,doublecopyzeros,exceptionalzeros}. It is therefore important to determine whether zeros and their reconstruction power reflect a more general principle or depend on this cubic origin. Pure $\tr(\phi^m)$ theories provide a direct way to test this question without introducing spin or derivative interactions.

\begin{figure*}[!t]
\centering
\input{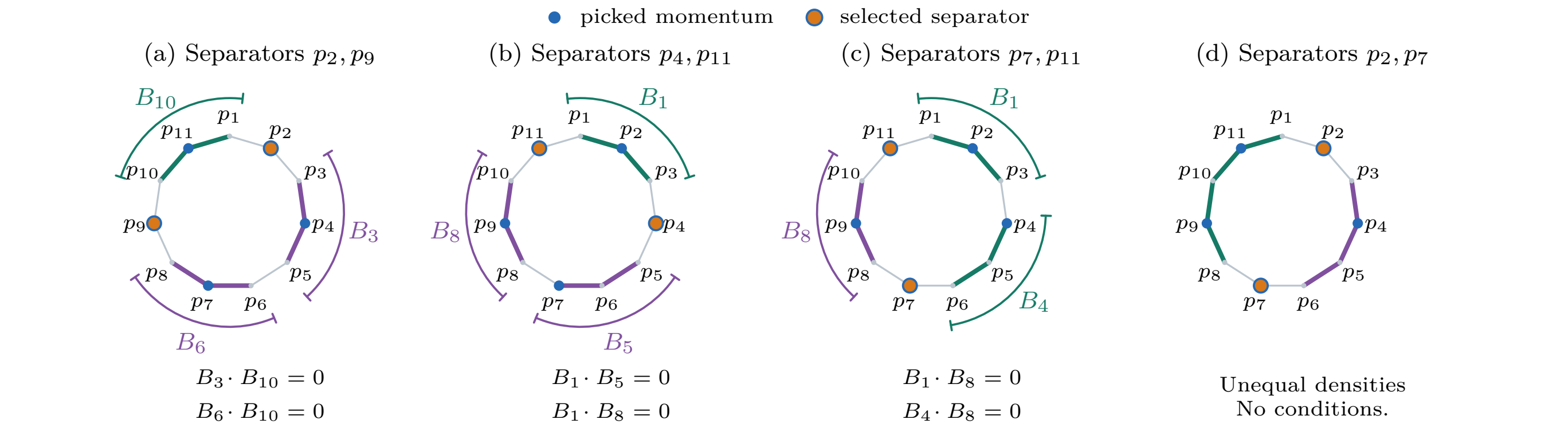}
\caption{The complete block zero in $\tr(\phi^5)$ with $V=3$ and $n=11$, defined by the $M=5$ picked momenta $p_2,p_4,p_7,p_9,p_{11}$. Vertices are momenta, with the initial selection in blue and the chosen separators in orange. The brackets identify the blocks $B_i$. Purple and green mark arcs $A$ and $B$. Their picked-momentum densities $(r_A/N_A,r_B/N_B)$ are $(2/6,1/3)$ in (a) and (b), $(1/3,2/6)$ in (c), and $(1/4,2/5)$ in (d), always excluding the separators. Of the ten candidate separator pairs, exactly the three in (a)--(c) have equal density. Impose every displayed equation from these panels together. The contraction $B_1\cdot B_8$ occurs twice, leaving five distinct conditions. Panel (d) has unequal densities and contributes no conditions. The other six discarded pairs are not drawn.}
\label{fig:equal-density-eleven}
\end{figure*}

Extending the rectangle zeros has nevertheless proved nontrivial, as the next natural target, $\tr(\phi^4)$, does not seem to admit a comparable general pattern of simple linear zeros.\footnote{For example, $\tr(\phi^4)$ amplitudes follow from a large shift of $\tr(\phi^3)$ channels with the same sign for the even--even and odd--odd channels~\cite{nlsmembedding}. Unlike the opposite-sign shift producing the NLSM, this shift does not preserve the cubic rectangle conditions.}

In this Letter we extend the zero structures and their reconstructive power to all $\tr(\phi^m)$. We find in fact that the simplest extension starts with $\tr(\phi^5)$ and continues to every odd $m$. The resulting \emph{block zeros} replace orthogonality between individual momenta by orthogonality between consecutive momentum blocks,
\begin{equation}
 B_i=p_i+\cdots+p_{i+m-3},\quad
 p_i\cdot p_j=0\ \longrightarrow\ B_i\cdot B_j=0.
 \label{eq:intro-blocks}
\end{equation}
A selection rule specifies which blocks must be orthogonal, recovering the usual rectangle zeros when $m=3$. We also find that these conditions recover locality and relative factorization residues in finite explicit checks, even when the ansatz allows crossing and repeated poles.

Returning to $\tr(\phi^4)$, we show that zeros can instead be specified by a nonlinear rank-one condition on a matrix of adjacent-pair contractions. Surpringly, this condition connects hidden zeros to threshold nullification, the classic cancellation of certain production amplitudes when all produced particles are at rest~\cite{voloshinloop,voloshinzeros,thresholdzeros}, and extends that zero to continuous real scattering configurations away from threshold. To our knowledge, threshold nullification has not been yet connected to hidden zero developments.

We then extend the matrix condition to a second family, the \emph{rod zeros}, described by a rod packing problem for every $m\geq4$. For this family, we prove that locality and the zero conditions together fix all relative diagram coefficients at every multiplicity.

\section{Block zeros for odd interactions}
\label{sec:odd-main}
We first introduce the block zeros, which generalize rectangular hidden zeros, which can be easily recovered when $m=3$.

We take massless outgoing momenta and define planar channels $X_{ab}=(p_a+\cdots+p_{b-1})^2$. A tree with $V$ interaction vertices has $n=(m-2)V+2$ external legs and $V-1$ propagators. Its diagrams correspond to divisions of an $n$-gon into $m$-gons by noncrossing diagonals, called $m$-angulations~\cite{abhy,stokes,accordiohedra,tropicalgeneral}.

For odd $m\geq3$ and $V\geq2$, the rule uses a set of nonadjacent external momenta. Schematically, the zero conditions have the form
\begin{equation}
 B_i\cdot B_j=0,\qquad B_i=\sum_{r=0}^{m-3}p_{i+r},
\end{equation}
where $B_i$ is a \emph{block momentum}. The rule below determines which blocks must be orthogonal.

We give the construction in four steps. To aid visualization, we display the momenta $p_i$ as the vertices of a polygon in cyclic order, as in Fig.~\ref{fig:equal-density-eleven}.

\paragraph{Step 1. Pick the momenta.} Pick
\begin{equation}
 M=\frac{m-3}{2}V+2
 \label{eq:density-count}
\end{equation}
nonadjacent external momenta. This choice defines the zero. The next steps determine all conditions defining this particular locus.

\paragraph{Step 2. Select the separator pairs.} For each picked pair $p_i,p_j$, let arc $A$ run from $p_{i+1}$ to $p_{j-1}$ and arc $B$ from $p_{j+1}$ to $p_{i-1}$, cyclically. Let $N_A,N_B$ count all momenta in the arcs and $r_A,r_B$ count the picked momenta among them. Retain the pair exactly when
\begin{equation}
 \frac{r_A}{N_A}=\frac{r_B}{N_B}.
 \label{eq:equal-density}
\end{equation}
The integer counts restrict the possible arc lengths. The common density is $(m-3)/[2(m-2)]$, so the arc lengths and picked counts must take the form
\begin{equation}
 \begin{aligned}
 N_A&=(m-2)h,&r_A&=\frac{m-3}{2}h,\\
 N_B&=(m-2)(V-h),&r_B&=\frac{m-3}{2}(V-h),
 \end{aligned}
 \label{eq:density-integer-counts}
\end{equation}
with $1\leq h<V$. Checking either arc suffices.\footnote{The density test keeps the locus larger. Imposing additional complete-block rectangles between picked separators still gives a zero, but restricts the locus further.}

\paragraph{Step 3. Impose the block rectangle.} Starting immediately after each separator, divide the arcs into consecutive blocks of $m-2$ momenta. Require every block on one arc to be orthogonal to every block on the other,
\begin{equation}
 \boxed{\begin{gathered}
 B_{i+1+\alpha(m-2)}\cdot B_{j+1+\beta(m-2)}=0,\\
 0\leq\alpha<h,\qquad0\leq\beta<V-h.
 \end{gathered}}
 \label{eq:density-rectangle}
\end{equation}
This is an $h\times(V-h)$ rectangle, of width $\min(h,V-h)$. 

\paragraph{Step 4. Combine the conditions.} For one fixed picked set, impose every retained rectangle simultaneously. Each picked set specifies one locus, and these loci are tested separately.\footnote{Picking every other position along a consecutive stretch, for example $\{p_n,p_2,p_4,\ldots,p_{n-V}\}$, selects width-one rectangles. This arrangement admits an optional relaxation to $B_c\cdot(\sum_{j\in I}B_{c+1+j(m-2)})=0$ for the prescribed consecutive groups $I$ in Sec.~\ref{sec:summed-strips}. Individual contractions may then be nonzero. At $m=3$, no blocks are combined.}

Every block equation is linear in physical channels, with the four-corner form in Eq.~\eqref{eq:blockmesh}. We prove
\begin{equation}
 A_n^{(m)}=0
 \label{eq:density-zero}
\end{equation}
on a nonempty regular locus for every picked set. Section~\ref{sec:density-proof} proves regularity and establishes vanishing by factorization. Section~\ref{sec:examples} gives higher-point examples in $\tr(\phi^5)$ and $\tr(\phi^7)$.

\subsection{The $\tr(\phi^3)$ limit and a $\tr(\phi^5)$ example}
For $m=3$, $M=2$ and exactly two nonadjacent momenta are picked. Neither arc contains another picked momentum, so $r_A=r_B=0$ and Eq.~\eqref{eq:equal-density} is automatic. Each block is simply one momentum, $B_i=p_i$, and the rule gives precisely
\begin{equation}
 p_a\cdot p_b=0,\qquad a\in A,\quad b\in B,
 \label{eq:density-cubic}
\end{equation}
the usual rectangle zeros~\cite{hidden}, with no additional conditions. At six points, picking $p_2,p_6$ gives the width-one zero
\begin{equation}
 p_1\cdot p_3=p_1\cdot p_4=p_1\cdot p_5=0,
 \label{eq:cubic-six-skinny}
\end{equation}
while picking $p_3,p_6$ gives the width-two zero
\begin{equation}
 p_a\cdot p_b=0,\qquad a=1,2,\quad b=4,5.
 \label{eq:cubic-six-wide}
\end{equation}
Thus the cubic specialization recovers the rectangular family at every multiplicity. 

For $\tr(\phi^5)$ at eight points, pick for example $p_2,p_4,p_6,p_8$. The only retained separator pairs are $(p_4,p_8)$ and $(p_2,p_6)$. Each intervening arc has three momenta, one of them picked. With $B_i=p_i+p_{i+1}+p_{i+2}$, their conditions are
\begin{equation}
 \begin{aligned}
 B_1\cdot B_5=0&\quad\Longleftrightarrow\quad X_{15}+X_{48}=0,\\
 B_3\cdot B_7=0&\quad\Longleftrightarrow\quad X_{26}+X_{37}=0.
 \end{aligned}
 \label{eq:quintic-eight-zero-main}
\end{equation}
They cancel the complete amplitude in pairs,
\begin{equation}
 A_8^{(5)}=
 \left(\frac1{X_{15}}+\frac1{X_{48}}\right)
 +\left(\frac1{X_{26}}+\frac1{X_{37}}\right)=0.
 \label{eq:quintic-eight-main}
\end{equation}
Figure~\ref{fig:equal-density-eleven} shows all three retained separator pairs for one picked set at eleven points, together with a discarded pair. The retained rectangles give five distinct block conditions, all displayed below the polygons. The eleven- and fourteen-point channel conditions are given in Secs.~\ref{sec:eleven-channels-supp} and \ref{sec:fourteen-channels-supp}.

\section{A rank condition in $\tr(\phi^4)$}
\label{sec:quartic-main}
We now consider $\tr(\phi^4)$ theory. In this case, we only find generic and natural zero loci that are nonlinear. At six points, the three diagrams give
\begin{equation}
 \begin{aligned}
 A_6^{(4)}&=\frac1{X_{14}}+\frac1{X_{25}}+\frac1{X_{36}}\\
 &=\frac{X_{14}X_{25}+X_{14}X_{36}+X_{25}X_{36}}
 {X_{14}X_{25}X_{36}}.
 \end{aligned}
 \label{eq:quartic-six}
\end{equation}
Remarkably, the numerator has a simple matrix interpretation. For any even $n\geq6$, form the matrix of contractions between adjacent momentum pairs,
\begin{equation}
 C_{ij}=2(p_{2i+1}+p_{2i+2})\cdot(p_{2j+2}+p_{2j+3}),
 \label{eq:C}
\end{equation}
with $i,j=0,\ldots,n/2-1$ and cyclic labels. At six points it is
\begin{equation}
 C=\begin{pmatrix}
 X_{14}&-X_{14}-X_{36}&X_{36}\\
 X_{25}&X_{36}&-X_{25}-X_{36}\\
 -X_{14}-X_{25}&X_{14}&X_{25}
 \end{pmatrix}.
 \label{eq:C-six}
\end{equation}
Every $2\times2$ minor is the numerator of Eq.~\eqref{eq:quartic-six}, up to sign. Thus the amplitude at six points vanishes precisely when $\rank C=1$, provided all physical channels are nonzero. We show that the same condition gives zeros at every even multiplicity,
\begin{equation}
 \boxed{\rank C=1\quad\Longrightarrow\quad A_n^{(4)}=0.}
 \label{eq:rankone}
\end{equation}
All $2\times2$ minors must vanish. The determinant alone is uninformative because momentum conservation already makes every row and column sum zero. 

For $V=3$, the eight-point amplitude uses a $4\times4$ matrix. Its upper-left minor gives
\begin{equation}
 \det\begin{pmatrix}
 X_{14}&X_{16}-X_{14}-X_{36}\\
 X_{25}&X_{36}
 \end{pmatrix}=0.
 \label{eq:quartic-eight-minor}
\end{equation}
All the other minors are imposed as well. The full matrix appears in Sec.~\ref{sec:four-matrix-example}, Eq.~\eqref{eq:C-eight-full}. This minor already explains the cancellation mechanism. It gives
\begin{equation}
 \frac1{X_{16}}\left(\frac1{X_{14}}+\frac1{X_{25}}+\frac1{X_{36}}\right)
 =\frac1{X_{14}X_{36}}.
 \label{eq:quartic-current-product}
\end{equation}
The rank condition simplifies every proper interval in this way, and the final tree sum vanishes. The rod packing construction in the next section extends this mechanism to general $m$.

\begin{figure*}[t]
\centering
\begingroup
\definecolor{rodblue}{HTML}{2569A7}
\definecolor{rodgrid}{HTML}{B4BDC5}
\begin{tikzpicture}[x=1cm,y=1cm,line join=round,every node/.style={font=\footnotesize,inner sep=0pt}]
\node[anchor=center,font=\fontsize{8.5}{10.5}\selectfont] at (0.15,0.12) {1};
\node[anchor=center,font=\fontsize{8.5}{10.5}\selectfont] at (0.44999999999999996,0.12) {2};
\node[anchor=center,font=\fontsize{8.5}{10.5}\selectfont] at (0.75,0.12) {3};
\node[anchor=center,font=\fontsize{8.5}{10.5}\selectfont] at (1.05,0.12) {4};
\node[anchor=center,font=\fontsize{8.5}{10.5}\selectfont] at (1.3499999999999999,0.12) {5};
\node[anchor=center,font=\fontsize{8.5}{10.5}\selectfont] at (1.65,0.12) {6};
\fill[rodblue!18] (0.0,-0.28) rectangle (0.6,0.0);
\fill[rodblue!18] (0.6,-0.28) rectangle (1.2,0.0);
\draw[rodgrid!60,line width=0.35pt] (0.0,-0.28)--(0.0,0.0);
\draw[rodgrid!60,line width=0.35pt] (0.3,-0.28)--(0.3,0.0);
\draw[rodgrid!60,line width=0.35pt] (0.6,-0.28)--(0.6,0.0);
\draw[rodgrid!60,line width=0.35pt] (0.8999999999999999,-0.28)--(0.8999999999999999,0.0);
\draw[rodgrid!60,line width=0.35pt] (1.2,-0.28)--(1.2,0.0);
\draw[rodgrid!60,line width=0.35pt] (1.5,-0.28)--(1.5,0.0);
\draw[rodgrid!60,line width=0.35pt] (1.7999999999999998,-0.28)--(1.7999999999999998,0.0);
\draw[rodgrid!60,line width=0.35pt] (0.0,-0.28)--(1.7999999999999998,-0.28);
\draw[rodgrid!60,line width=0.35pt] (0.0,0.0)--(1.7999999999999998,0.0);
\draw[rodblue,line width=.85pt] (0.0,-0.28) rectangle (0.6,0.0);
\draw[rodblue,line width=.85pt] (0.6,-0.28) rectangle (1.2,0.0);
\node[anchor=center,font=\fontsize{8.5}{10.5}\selectfont] at (0.9,-0.57) {$(X_{1,4}X_{3,6})^{-1}$};
\node[anchor=center,font=\fontsize{9}{11}\selectfont] at (1.975,-0.57) {$+$};
\node[anchor=center,font=\fontsize{8.5}{10.5}\selectfont] at (2.3,0.12) {1};
\node[anchor=center,font=\fontsize{8.5}{10.5}\selectfont] at (2.5999999999999996,0.12) {2};
\node[anchor=center,font=\fontsize{8.5}{10.5}\selectfont] at (2.9,0.12) {3};
\node[anchor=center,font=\fontsize{8.5}{10.5}\selectfont] at (3.2,0.12) {4};
\node[anchor=center,font=\fontsize{8.5}{10.5}\selectfont] at (3.5,0.12) {5};
\node[anchor=center,font=\fontsize{8.5}{10.5}\selectfont] at (3.8,0.12) {6};
\fill[rodblue!18] (2.15,-0.28) rectangle (2.75,0.0);
\fill[rodblue!18] (3.05,-0.28) rectangle (3.65,0.0);
\draw[rodgrid!60,line width=0.35pt] (2.15,-0.28)--(2.15,0.0);
\draw[rodgrid!60,line width=0.35pt] (2.4499999999999997,-0.28)--(2.4499999999999997,0.0);
\draw[rodgrid!60,line width=0.35pt] (2.75,-0.28)--(2.75,0.0);
\draw[rodgrid!60,line width=0.35pt] (3.05,-0.28)--(3.05,0.0);
\draw[rodgrid!60,line width=0.35pt] (3.3499999999999996,-0.28)--(3.3499999999999996,0.0);
\draw[rodgrid!60,line width=0.35pt] (3.65,-0.28)--(3.65,0.0);
\draw[rodgrid!60,line width=0.35pt] (3.9499999999999997,-0.28)--(3.9499999999999997,0.0);
\draw[rodgrid!60,line width=0.35pt] (2.15,-0.28)--(3.9499999999999997,-0.28);
\draw[rodgrid!60,line width=0.35pt] (2.15,0.0)--(3.9499999999999997,0.0);
\draw[rodblue,line width=.85pt] (2.15,-0.28) rectangle (2.75,0.0);
\draw[rodblue,line width=.85pt] (3.05,-0.28) rectangle (3.65,0.0);
\node[anchor=center,font=\fontsize{8.5}{10.5}\selectfont] at (3.05,-0.57) {$(X_{1,4}X_{4,7})^{-1}$};
\node[anchor=center,font=\fontsize{9}{11}\selectfont] at (4.125,-0.57) {$+$};
\node[anchor=center,font=\fontsize{8.5}{10.5}\selectfont] at (4.45,0.12) {1};
\node[anchor=center,font=\fontsize{8.5}{10.5}\selectfont] at (4.75,0.12) {2};
\node[anchor=center,font=\fontsize{8.5}{10.5}\selectfont] at (5.05,0.12) {3};
\node[anchor=center,font=\fontsize{8.5}{10.5}\selectfont] at (5.35,0.12) {4};
\node[anchor=center,font=\fontsize{8.5}{10.5}\selectfont] at (5.6499999999999995,0.12) {5};
\node[anchor=center,font=\fontsize{8.5}{10.5}\selectfont] at (5.949999999999999,0.12) {6};
\fill[rodblue!18] (4.3,-0.28) rectangle (4.8999999999999995,0.0);
\fill[rodblue!18] (5.5,-0.28) rectangle (6.1,0.0);
\draw[rodgrid!60,line width=0.35pt] (4.3,-0.28)--(4.3,0.0);
\draw[rodgrid!60,line width=0.35pt] (4.6,-0.28)--(4.6,0.0);
\draw[rodgrid!60,line width=0.35pt] (4.8999999999999995,-0.28)--(4.8999999999999995,0.0);
\draw[rodgrid!60,line width=0.35pt] (5.199999999999999,-0.28)--(5.199999999999999,0.0);
\draw[rodgrid!60,line width=0.35pt] (5.5,-0.28)--(5.5,0.0);
\draw[rodgrid!60,line width=0.35pt] (5.8,-0.28)--(5.8,0.0);
\draw[rodgrid!60,line width=0.35pt] (6.1,-0.28)--(6.1,0.0);
\draw[rodgrid!60,line width=0.35pt] (4.3,-0.28)--(6.1,-0.28);
\draw[rodgrid!60,line width=0.35pt] (4.3,0.0)--(6.1,0.0);
\draw[rodblue,line width=.85pt] (4.3,-0.28) rectangle (4.8999999999999995,0.0);
\draw[rodblue,line width=.85pt] (5.5,-0.28) rectangle (6.1,0.0);
\node[anchor=center,font=\fontsize{8.5}{10.5}\selectfont] at (5.2,-0.57) {$(X_{1,4}X_{5,8})^{-1}$};
\node[anchor=center,font=\fontsize{9}{11}\selectfont] at (6.275,-0.57) {$+$};
\node[anchor=center,font=\fontsize{8.5}{10.5}\selectfont] at (6.6,0.12) {1};
\node[anchor=center,font=\fontsize{8.5}{10.5}\selectfont] at (6.8999999999999995,0.12) {2};
\node[anchor=center,font=\fontsize{8.5}{10.5}\selectfont] at (7.199999999999999,0.12) {3};
\node[anchor=center,font=\fontsize{8.5}{10.5}\selectfont] at (7.499999999999999,0.12) {4};
\node[anchor=center,font=\fontsize{8.5}{10.5}\selectfont] at (7.799999999999999,0.12) {5};
\node[anchor=center,font=\fontsize{8.5}{10.5}\selectfont] at (8.1,0.12) {6};
\fill[rodblue!18] (6.749999999999999,-0.28) rectangle (7.349999999999999,0.0);
\fill[rodblue!18] (7.35,-0.28) rectangle (7.949999999999999,0.0);
\draw[rodgrid!60,line width=0.35pt] (6.449999999999999,-0.28)--(6.449999999999999,0.0);
\draw[rodgrid!60,line width=0.35pt] (6.749999999999999,-0.28)--(6.749999999999999,0.0);
\draw[rodgrid!60,line width=0.35pt] (7.049999999999999,-0.28)--(7.049999999999999,0.0);
\draw[rodgrid!60,line width=0.35pt] (7.35,-0.28)--(7.35,0.0);
\draw[rodgrid!60,line width=0.35pt] (7.6499999999999995,-0.28)--(7.6499999999999995,0.0);
\draw[rodgrid!60,line width=0.35pt] (7.949999999999999,-0.28)--(7.949999999999999,0.0);
\draw[rodgrid!60,line width=0.35pt] (8.25,-0.28)--(8.25,0.0);
\draw[rodgrid!60,line width=0.35pt] (6.449999999999999,-0.28)--(8.25,-0.28);
\draw[rodgrid!60,line width=0.35pt] (6.449999999999999,0.0)--(8.25,0.0);
\draw[rodblue,line width=.85pt] (6.749999999999999,-0.28) rectangle (7.349999999999999,0.0);
\draw[rodblue,line width=.85pt] (7.35,-0.28) rectangle (7.949999999999999,0.0);
\node[anchor=center,font=\fontsize{8.5}{10.5}\selectfont] at (7.35,-0.57) {$(X_{2,5}X_{4,7})^{-1}$};
\node[anchor=center,font=\fontsize{9}{11}\selectfont] at (8.424999999999999,-0.57) {$+$};
\node[anchor=center,font=\fontsize{8.5}{10.5}\selectfont] at (8.75,0.12) {1};
\node[anchor=center,font=\fontsize{8.5}{10.5}\selectfont] at (9.049999999999999,0.12) {2};
\node[anchor=center,font=\fontsize{8.5}{10.5}\selectfont] at (9.35,0.12) {3};
\node[anchor=center,font=\fontsize{8.5}{10.5}\selectfont] at (9.65,0.12) {4};
\node[anchor=center,font=\fontsize{8.5}{10.5}\selectfont] at (9.95,0.12) {5};
\node[anchor=center,font=\fontsize{8.5}{10.5}\selectfont] at (10.25,0.12) {6};
\fill[rodblue!18] (8.9,-0.28) rectangle (9.5,0.0);
\fill[rodblue!18] (9.799999999999999,-0.28) rectangle (10.399999999999999,0.0);
\draw[rodgrid!60,line width=0.35pt] (8.6,-0.28)--(8.6,0.0);
\draw[rodgrid!60,line width=0.35pt] (8.9,-0.28)--(8.9,0.0);
\draw[rodgrid!60,line width=0.35pt] (9.2,-0.28)--(9.2,0.0);
\draw[rodgrid!60,line width=0.35pt] (9.5,-0.28)--(9.5,0.0);
\draw[rodgrid!60,line width=0.35pt] (9.799999999999999,-0.28)--(9.799999999999999,0.0);
\draw[rodgrid!60,line width=0.35pt] (10.1,-0.28)--(10.1,0.0);
\draw[rodgrid!60,line width=0.35pt] (10.399999999999999,-0.28)--(10.399999999999999,0.0);
\draw[rodgrid!60,line width=0.35pt] (8.6,-0.28)--(10.399999999999999,-0.28);
\draw[rodgrid!60,line width=0.35pt] (8.6,0.0)--(10.399999999999999,0.0);
\draw[rodblue,line width=.85pt] (8.9,-0.28) rectangle (9.5,0.0);
\draw[rodblue,line width=.85pt] (9.799999999999999,-0.28) rectangle (10.399999999999999,0.0);
\node[anchor=center,font=\fontsize{8.5}{10.5}\selectfont] at (9.5,-0.57) {$(X_{2,5}X_{5,8})^{-1}$};
\node[anchor=center,font=\fontsize{9}{11}\selectfont] at (10.575,-0.57) {$+$};
\node[anchor=center,font=\fontsize{8.5}{10.5}\selectfont] at (10.9,0.12) {1};
\node[anchor=center,font=\fontsize{8.5}{10.5}\selectfont] at (11.2,0.12) {2};
\node[anchor=center,font=\fontsize{8.5}{10.5}\selectfont] at (11.5,0.12) {3};
\node[anchor=center,font=\fontsize{8.5}{10.5}\selectfont] at (11.8,0.12) {4};
\node[anchor=center,font=\fontsize{8.5}{10.5}\selectfont] at (12.1,0.12) {5};
\node[anchor=center,font=\fontsize{8.5}{10.5}\selectfont] at (12.4,0.12) {6};
\fill[rodblue!18] (11.35,-0.28) rectangle (11.95,0.0);
\fill[rodblue!18] (11.95,-0.28) rectangle (12.549999999999999,0.0);
\draw[rodgrid!60,line width=0.35pt] (10.75,-0.28)--(10.75,0.0);
\draw[rodgrid!60,line width=0.35pt] (11.05,-0.28)--(11.05,0.0);
\draw[rodgrid!60,line width=0.35pt] (11.35,-0.28)--(11.35,0.0);
\draw[rodgrid!60,line width=0.35pt] (11.65,-0.28)--(11.65,0.0);
\draw[rodgrid!60,line width=0.35pt] (11.95,-0.28)--(11.95,0.0);
\draw[rodgrid!60,line width=0.35pt] (12.25,-0.28)--(12.25,0.0);
\draw[rodgrid!60,line width=0.35pt] (12.55,-0.28)--(12.55,0.0);
\draw[rodgrid!60,line width=0.35pt] (10.75,-0.28)--(12.55,-0.28);
\draw[rodgrid!60,line width=0.35pt] (10.75,0.0)--(12.55,0.0);
\draw[rodblue,line width=.85pt] (11.35,-0.28) rectangle (11.95,0.0);
\draw[rodblue,line width=.85pt] (11.95,-0.28) rectangle (12.549999999999999,0.0);
\node[anchor=center,font=\fontsize{8.5}{10.5}\selectfont] at (11.65,-0.57) {$(X_{3,6}X_{5,8})^{-1}$};
\node[anchor=east,font=\fontsize{9.5}{11.5}\selectfont] at (-0.35,-0.57) {$S_2(1,5)={}$};
\end{tikzpicture}
\endgroup
\caption{The packing sum $S_2(1,5)$ for $\tr(\phi^4)$ at eight points. Two rods of length two occupy six cells, with allowed starting positions $1,\ldots,5$. Each rod starting at $i$ has weight $1/X_{i,i+3}$. Multiply the two rod weights within each packing and add the six resulting terms. In general, $k$ counts the rods and $a,b$ bound their starting positions in $S_k(a,b)$.}
\label{fig:rod-main}
\end{figure*}

\subsection{Real scattering and threshold nullification}
\label{sec:physical-four}
We connect the $\tr(\phi^4)$ rank condition to threshold nullification. In massive unbroken $\phi^4$ theory, the full tree amplitude for $2\to r$ scattering vanishes for every even $r>4$ when all outgoing particles are at rest, through cancellations among finite diagrams~\cite{voloshinloop,voloshinzeros,thresholdzeros}. We show that the corresponding ordered zero belongs to a continuous family of real zeros extending above threshold.

Real channel values need not come from on-shell Lorentzian momenta satisfying momentum conservation and the Gram constraints. On the rank-one locus, however, we show that every regular real point admits such a realization in $1+2$ dimensions with a common mass. Section~\ref{sec:real-rank} gives the construction, with alternating incoming and outgoing legs describing $n/2\to n/2$ scattering.

To reach threshold production, consider instead two incoming particles. Let $n=2V+2$ and give all external and exchanged scalars mass $M\geq0$, with inverse propagators $\widehat X_{ab}=(p_a+\cdots+p_{b-1})^2-M^2$. In mostly-minus signature, take adjacent back-to-back pairs,
\begin{equation}
 \begin{gathered}
 p_{2i-1}=(b_i/2,\mathbf q_i),\qquad
 p_{2i}=(b_i/2,-\mathbf q_i),\\
 |\mathbf q_i|^2=b_i^2/4-M^2,\qquad
 \sum_{i=1}^{V+1}b_i=0,
 \end{gathered}
 \label{eq:physical-pairs}
\end{equation}
with $|b_i|\geq2M$. Each pair carries only energy, so the matrix in Eq.~\eqref{eq:C} factorizes as
\begin{equation}
 C_{i-1,j-1}=b_i(b_j+b_{j+1}).
 \label{eq:physical-rank}
\end{equation}
For regular configurations this has rank one, independently of the pair directions.

An allowed interval contains complete pairs with total energy $B$ and one constituent of a pair with energy $b$. Its inverse propagator is
\begin{equation}
 \begin{aligned}
 \widehat X&=(B+b/2)^2-(b^2/4-M^2)-M^2\\
 &=B(B+b).
 \end{aligned}
 \label{eq:physical-mass-cancel}
\end{equation}
At fixed pair energies, the channels are independent of the pair directions and mass. The rank-one proof applies to $\widehat X$ because its adjacent boundary entries vanish and its cyclic mixed differences still give $C$.

Taking $b_1<0$ and all other $b_i>0$ gives $2\to n-2$ scattering with legs $1,2$ incoming. Each factor $B$ or $B+b$ is a nonempty proper subset sum of pair energies. It is positive if it excludes $b_1$, and otherwise equals minus the positive complementary sum. Thus no physical propagator is singular.

For $M>0$, the threshold configuration is
\begin{equation}
 b_1=-2VM,\qquad b_2=\cdots=b_{V+1}=2M.
 \label{eq:physical-threshold}
\end{equation}
All $2V$ outgoing particles are then at rest. Increasing their pair energies and adjusting $b_1$ to conserve momentum moves above threshold while preserving the regular zero. The ordered amplitude with adjacent incoming legs already vanishes for $2\to4$, whereas the full single-scalar amplitude first vanishes for $2\to6$. Away from threshold, the ordered zero need not survive summing over orderings.

We thus connect threshold nullification to hidden zeros. The ordered threshold zero belongs to a continuous family of real $2\to n-2$ scattering zeros on the rank-one locus, extending above threshold with every propagator finite.

\section{Rod packing zeros for any $\tr(\phi^m)$}
\label{sec:product-main}
The eight-point cancellation in $\tr(\phi^4)$ suggests interpreting products of inverse channels as weights of nonoverlapping rods. Equation~\eqref{eq:quartic-current-product} then gives $X_{16}$ as the sum of single-rod weights divided by the complete-tiling weight. We extend this packing rule to every $m\geq4$ and allowed multiplicity.

For $n=(m-2)V+2$ with $V\geq2$, choose a root leg and relabel it $n$, opening the remaining ordering into a line of $n-2$ gap cells between legs $1,\ldots,n-1$. A rod now covers $m-2$ consecutive cells and carries weight $1/X_{i,i+m-1}$ when it starts at cell $i$. Rods may touch but cannot overlap, and multiplying their weights gives the weight of a packing.

Let $S_k(a,b)$ be the sum of the weights of all packings of $k$ rods whose starting cells lie between $a$ and $b$,
\begin{equation}
 S_k(a,b)=
 \sum_{\substack{a\leq i_1<\cdots<i_k\leq b\\i_{r+1}-i_r\geq m-2}}\ 
 \prod_{r=1}^{k}\frac1{X_{i_r,i_r+m-1}}.
 \label{eq:rod-sum-definition}
\end{equation}
The empty packing has weight $S_0(a,b)=1$, while the sum is zero if no packing with $k>0$ rods fits.

We define the rod-zero family $\G$ by two channel assignments. For every interval with $1\leq a<b\leq n$, $b-a=(m-2)\ell+1$, and $2\leq\ell<V$, set
\begin{equation}
 \boxed{X_{ab}=\frac{S_{\ell-1}(a,b-m+1)}{S_\ell(a,b-m+1)}.}
 \label{eq:closed-long-main}
\end{equation}
For the rod covering the last $m-2$ cells, set
\begin{equation}
 \boxed{X_{n-m+1,n}=-
 \frac{S_{V-2}(1,n-2m+3)}{S_{V-1}(1,n-m)}.}
 \label{eq:closed-main}
\end{equation}
At $V=2$, only the second assignment is needed. All physical channels and displayed denominators must be nonzero.

The family has $n-m$ free parameters in the space of $n(V-1)/2$ physical channels, and hence
\begin{equation}
 \#\text{ independent conditions}=\frac{n(V-1)}2-(n-m).
 \label{eq:rod-condition-count}
\end{equation}
The $n$ choices of root give distinct loci for $m\geq5,V\geq3$, but the same locus when $m=4$ or $V=2$ (Sec.~\ref{sec:rod-root-count}). Section~\ref{sec:product} proves vanishing and regularity at every allowed multiplicity.

\paragraph{Eight-point example.}
For $\tr(\phi^4)$ at $n=8$, the interval assignment gives
\begin{equation}
 X_{16}=X_{14}X_{36}\left(\frac1{X_{14}}+\frac1{X_{25}}+\frac1{X_{36}}\right).
 \label{eq:quartic-five-leg-rods}
\end{equation}
Shifting the labels by one and two gives $X_{27}$ and $X_{38}$. The remaining assignment is
\begin{equation}
 X_{58}=-\frac{\displaystyle\frac1{X_{14}}+\frac1{X_{25}}+\frac1{X_{36}}}
 {\displaystyle\frac1{X_{14}X_{36}}+\frac1{X_{14}X_{47}}+\frac1{X_{25}X_{47}}}.
 \label{eq:quartic-eight-rods}
\end{equation}
These four relations reduce the twelve diagrams to the six packing weights in Fig.~\ref{fig:rod-main}, whose sum vanishes. The $X_{16}$ relation is the vanishing minor in Eq.~\eqref{eq:quartic-eight-minor}, and all four relations together are equivalent to $\rank C=1$ on the regular locus.

Section~\ref{sec:six-rods} gives explicit $\tr(\phi^6)$ solutions at ten and fourteen points.

\section{Locality and unitarity from zeros}
\label{sec:unique-main}
We now discuss the reconstruction of amplitudes from trial expressions by imposing the two types of zero conditions.

\subsection{Locality and unitarity from block zeros}
To test whether zeros determine locality, take $\widetilde A_n$ to be a sum of every product of $V-1$ inverse physical channels with independent constant coefficients, including crossing and repeated poles. No cyclic symmetry is assumed. We impose every rectangle selected by each tested picked set, treating its locus separately. In every case in Table~\ref{tab:matching-ranks}, only $\widetilde A_n=cA_n^{(m)}$ survives. These finite results recover pole compatibility and relative factorization residues from the zeros and ansatz assumptions. We conjecture that the general rule does so at every odd $m$ and allowed multiplicity.

\subsection{Unitarity from rod zeros and locality}
With locality assumed, we take an ansatz with constant coefficients
\begin{equation}
 \widetilde A_n=\sum_{D\in\calD_{n,m}}c_D\prod_{e\in D}\frac1{X_e},
 \label{eq:localansatz}
\end{equation}
where $D$ runs over compatible planar trees with vertices of valence $m$.

For rod zeros, we prove uniqueness with locality for every $m\geq4$ and every allowed $n$. Limits within $\G$ isolate an end vertex and leave a freely varying rod zero at lower multiplicity. Induction fixes the coefficients of diagrams containing that vertex. Diagrams sharing end vertices connect these equalities, forcing all $c_D$ to agree. Section~\ref{sec:uniqueness} gives the proof.

The zero conditions fix $\widetilde A_n=c_VA_n^{(m)}$ but leave one normalization $c_V$ for each number of vertices. Factorization into separately normalized lower amplitudes requires $c_V=c_{V_L}c_{V_R}$ whenever $V_L+V_R=V$. Choosing $c_V=g^V$, with elementary coupling $g$, satisfies this condition. The homogeneous zero tests fix relative residues but do not relate these overall factors across multiplicities.
\begin{table}[t]
\caption{Largest completed checks for each reported interaction order.  Each check mark denotes uniqueness up to normalization without locality assumed. No counterexample was found in these checks.}
\label{tab:matching-ranks}
\setlength{\tabcolsep}{3pt}
\begin{ruledtabular}
\begin{tabular}{rrrrrc}
$m$&$V$&$n$&\shortstack{Zero loci\\used}&\shortstack{Unrestricted\\terms}&\shortstack{Locality and\\unitarity emerge}\\\hline
5&7&23&46&$185\,250\,786$&$\checkmark$\\
7&6&32&32&$30\,872\,016$&$\checkmark$\\
9&6&44&44&$146\,803\,272$&$\checkmark$\\
21&5&97&97&$60\,862\,165$&$\checkmark$\\
101&4&398&398&$35\,641\,099$&$\checkmark$\\
\end{tabular}
\end{ruledtabular}
\end{table}
\section{Discussion}
\label{sec:discussion-main}
Regular kinematic cancellations constrain diagram weights and allowed pole combinations beyond $\tr(\phi^3)$. We establish block zeros through the equal-density rule at every odd valence and allowed multiplicity. Their reconstruction power is supported by the exact checks. For every $m\geq4$, we prove that rod zeros determine all relative diagram coefficients when locality is assumed. The $\tr(\phi^4)$ rank condition also connects these algebraic cancellations to real scattering and threshold nullification.

The main open question is to derive both locality and factorization from zeros at every interaction order and multiplicity. The optional sums of block contractions in Sec.~\ref{sec:summed-strips} may be evidence of further structure. An equations-of-motion derivation based on the methods of Ref.~\cite{classical} could explain these zeros or reveal new ones. The structure of these summed conditions seems particularly natural in that approach. Finally, threshold nullification also occurs in $\phi^3$ theory, where it appears to be connected to the continuous $\tr(\phi^3)$ zeros~\cite{splitting}.

\medskip
\noindent\emph{Acknowledgments.} LR is supported by the National Natural Science Foundation of China General Program No.~12475070 and the Beijing Natural Science Foundation International Scientist Project No.~IS24014. AI tools were used to assist with checks and editing.

\bibliography{references}

\clearpage
\onecolumngrid
\renewcommand{\theHequation}{supp.\arabic{equation}}
\renewcommand{\theHtable}{supp.\arabic{table}}
\renewcommand{\theHfigure}{supp.\arabic{figure}}
\setcounter{equation}{0}
\setcounter{table}{0}
\setcounter{figure}{0}
\setcounter{section}{0}
\setcounter{subsection}{0}
\setcounter{secnumdepth}{2}
\renewcommand{\thesection}{S\arabic{section}}
\renewcommand{\thesubsection}{\Alph{subsection}}
\renewcommand{\theHsection}{supp.\arabic{section}}
\renewcommand{\theHsubsection}{supp.\arabic{section}.\arabic{subsection}}
\renewcommand{\theequation}{S\arabic{equation}}
\renewcommand{\thetable}{S\arabic{table}}
\renewcommand{\thefigure}{S\arabic{figure}}
\begin{center}
{\large\bf Supplemental Material}\par\medskip
{\large Beyond hidden zeros and uniqueness in $\tr(\phi^3)$: Block and rod zeros in any $\tr(\phi^m)$}
\end{center}

We first give explicit examples, then prove the block and rod zeros and the uniqueness result for rod zeros with locality. The last section gives the real-momentum construction and further threshold connections. Throughout the algebraic arguments, the physical channels are independent, $n=(m-2)V+2$, and the vertices are normalized to one. Regularity means that every physical channel is nonzero. Only the last section imposes a fixed-dimensional momentum realization.

\section{Explicit examples of zero loci}
\label{sec:examples}
All examples describe loci in physical channels. Every physical channel and displayed denominator must be nonzero.

\label{sec:boundary-mesh}
An elementary mesh cell carries $c_{ij}=-2p_i\cdot p_j$. Summing over two blocks of $m-2$ momenta cancels the interior channels and leaves four corners,
\begin{equation}
 \begin{aligned}
 -2B_i\cdot B_j&=\sum_{r,s=0}^{m-3}c_{i+r,j+s}\\
 &=X_{ij}+X_{i+m-2,j+m-2}-X_{i,j+m-2}-X_{i+m-2,j}.
 \end{aligned}
 \label{eq:blockmesh}
\end{equation}
A retained separator pair selects an $h\times(V-h)$ rectangle of these tiles. Each tile total vanishes separately. The elementary contractions inside one tile need not vanish.

\subsection{The eleven-point example in $\tr(\phi^5)$}
\label{sec:eleven-channels-supp}
Pick $p_2,p_4,p_7,p_9,p_{11}$. The retained separator pairs are $(2,9)$, $(4,11)$, and $(7,11)$, each with a $1\times2$ rectangle. Their six drawn tiles include one repeated contraction and give
\begin{equation}
 \begin{gathered}
 B_1\cdot B_5=B_1\cdot B_8=B_4\cdot B_8=0,\\
 B_{10}\cdot B_3=B_{10}\cdot B_6=0.
 \end{gathered}
 \label{eq:quintic-eleven-blocks}
\end{equation}
The five independent channel relations are
\begin{equation}
 \begin{gathered}
 X_{18}=X_{15}+X_{48},\qquad X_{4,11}=-X_{15},\\
 X_{7,11}=-X_{18},\qquad X_{3,10}=-X_{29},\\
 X_{6,10}=X_{26}-X_{29}.
 \end{gathered}
 \label{eq:q11}
\end{equation}
They leave six physical channels unfixed and make the complete amplitude vanish throughout the regular locus.

\subsection{The fourteen-point example in $\tr(\phi^5)$}
\label{sec:fourteen-channels-supp}
At fourteen points, pick $p_2,p_4,p_7,p_{10},p_{12},p_{14}$. The retained separator pairs are $(2,12)$, $(4,14)$, $(7,14)$, and $(10,14)$. These four rectangles give nine distinct block conditions. The pair $(7,14)$ contributes the $2\times2$ rectangle with rows $B_1,B_4$ and columns $B_8,B_{11}$, while the other three have width one. With $B_i=p_i+p_{i+1}+p_{i+2}$, impose
\begin{equation}
 \begin{gathered}
 B_1\cdot B_5=B_1\cdot B_8=B_1\cdot B_{11}=0,\\
 B_4\cdot B_8=B_4\cdot B_{11}=B_7\cdot B_{11}=0,\\
 B_{13}\cdot B_3=B_{13}\cdot B_6=B_{13}\cdot B_9=0.
 \end{gathered}
\end{equation}

The equivalent channel system is
\begin{equation}
 \begin{aligned}
 X_{18}&=X_{15}+X_{48},&
 X_{1,11}&=X_{18}+X_{7,11},&
 X_{4,11}&=X_{48}+X_{7,11},\\
 X_{4,14}&=-X_{15},&
 X_{7,14}&=-X_{18},&
 X_{10,14}&=-X_{1,11},\\
 X_{3,13}&=-X_{2,12},&
 X_{6,13}&=X_{26}-X_{2,12},&
 X_{9,13}&=X_{29}-X_{2,12}.
 \end{aligned}
 \label{eq:phi5-fourteen-channels}
\end{equation}
The pairs $(4,14)$, $(7,14)$, and $(10,14)$ give the first two lines, while $(2,12)$ gives the last. Nine independent equations in 21 physical channels leave twelve free parameters.

\subsection{The seventeen-point example in $\tr(\phi^7)$}
\label{sec:seven-example}
Here $V=3$ and every block contains five momenta. Pick
\begin{equation}
 \{p_2,p_4,p_6,p_8,p_{11},p_{13},p_{15},p_{17}\}.
\end{equation}
The retained separator pairs are $(2,8)$, $(2,13)$, $(4,15)$, $(6,17)$, and $(11,17)$. Their $1\times2$ rectangles give eight distinct conditions,
\begin{equation}
 \begin{gathered}
 B_1\cdot B_7=B_1\cdot B_{12}=B_6\cdot B_{12}=0,\\
 B_3\cdot B_9=B_3\cdot B_{14}=B_8\cdot B_{14}=0,\\
 B_{16}\cdot B_5=B_{16}\cdot B_{10}=0.
 \end{gathered}
\end{equation}
In physical channels they are
\begin{equation}
 \begin{aligned}
 X_{1,12}&=X_{1,7}+X_{6,12},&X_{6,17}&=-X_{1,7},&X_{11,17}&=-X_{1,12},\\
 X_{3,14}&=X_{3,9}+X_{8,14},&X_{2,8}&=-X_{3,9},&X_{2,13}&=-X_{3,14},\\
 X_{5,16}&=-X_{4,15},&X_{10,16}&=X_{4,10}-X_{4,15}.&&
 \end{aligned}
 \label{eq:seven-seventeen-channels}
\end{equation}
Nine physical channels remain unfixed. These relations make all 51 diagrams cancel on the regular locus.

\subsection{The full matrix in $\tr(\phi^4)$ at eight points}
\label{sec:four-matrix-example}
All eight physical channels enter the matrix of adjacent-pair contractions,
\begin{equation}
 C=\begin{pmatrix}
 X_{14}&X_{16}-X_{14}-X_{36}&X_{36}-X_{16}-X_{38}&X_{38}\\
 X_{25}&X_{36}&X_{38}-X_{36}-X_{58}&X_{58}-X_{38}-X_{25}\\
 X_{27}-X_{25}-X_{47}&X_{47}&X_{58}&X_{25}-X_{58}-X_{27}\\
 X_{47}-X_{27}-X_{14}&X_{14}-X_{47}-X_{16}&X_{16}&X_{27}
 \end{pmatrix}.
 \label{eq:C-eight-full}
\end{equation}
The zero requires every $2\times2$ minor to vanish, with every physical channel nonzero. For example, the upper-left minor gives $X_{25}(X_{16}-X_{14}-X_{36})=X_{14}X_{36}$, precisely the rod relation for $X_{16}$. Its shifts and the closing equation give the full rank-one condition, as in the main text.

\subsection{Explicit rod solutions in $\tr(\phi^6)$}
\label{sec:six-rods}
At ten points, $V=2$ and one rod of length four has five positions on eight cells. Leave $X_{1,6},X_{2,7},X_{3,8},X_{4,9}$ unfixed. The complete condition is
\begin{equation}
 X_{5,10}=-\left(\frac1{X_{1,6}}+\frac1{X_{2,7}}+\frac1{X_{3,8}}+\frac1{X_{4,9}}\right)^{-1}.
 \label{eq:six-ten-rods}
\end{equation}
There are no longer proper channels. The five single-rod weights sum to zero.

At fourteen points, $V=3$. Leave the eight shortest channels $X_{i,i+5}$, $i=1,\ldots,8$, unfixed. Two rods fit on twelve cells. Ten packings omit the last rod and five contain it, giving
\begin{equation}
 X_{9,14}=-\frac{\displaystyle\sum_{i=1}^{5}\frac1{X_{i,i+5}}}
 {\displaystyle\sum_{i=1}^{4}\sum_{j=i+4}^{8}\frac1{X_{i,i+5}X_{j,j+5}}}.
 \label{eq:six-fourteen-close}
\end{equation}
The other five channels are determined by the one-rod packings on eight cells,
\begin{equation}
 \begin{aligned}
 X_{a,a+9}=X_{a,a+5}X_{a+4,a+9}
 \biggl(&\frac1{X_{a,a+5}}+\frac1{X_{a+1,a+6}}+\frac1{X_{a+2,a+7}}\\
 &+\frac1{X_{a+3,a+8}}+\frac1{X_{a+4,a+9}}\biggr),
 \qquad a=1,\ldots,5.
 \end{aligned}
 \label{eq:six-fourteen-long}
\end{equation}
Together these equations give all fourteen physical channels in terms of the eight left unfixed. On the five longer-channel relations, the 35 diagrams reduce to fifteen packing weights,
\begin{equation}
 A_{14}^{(6)}=\sum_{i=1}^{5}\sum_{j=i+4}^{9}\frac1{X_{i,i+5}X_{j,j+5}},
\end{equation}
whose sum vanishes by Eq.~\eqref{eq:six-fourteen-close}. All physical channels and displayed denominators must remain nonzero in both examples.

\section{Block zeros for odd interactions}
\label{sec:odd}
\label{sec:density-proof}
We prove that every picked set in the equal-density prescription gives a regular zero at every allowed multiplicity. For $m=3$, the prescription is exactly the usual rectangle zero~\cite{hidden}. We therefore take odd $m\geq5$ below and write $q=(m-3)/2$. A zero at $V$ interaction vertices requires $qV+2$ nonadjacent picked momenta.

The proof uses induction on $V$. We first show that the locus has regular points. We then examine each possible pole of the amplitude restricted to that locus. Usually, one of its two factorization amplitudes inherits a lower-point zero. The remaining poles occur in crossing pairs whose residues cancel. Once all poles have canceled, homogeneity forces the amplitude to vanish. The pole limits are used only in this proof. The zero itself is defined with every propagator nonzero.

\subsection{Regularity and possible coincident poles}
For this argument, place $p_i$ on edge $(i,i+1)$ of the channel polygon and mark that edge when $p_i$ is picked. This is the polygon whose diagonals carry the $X_{ab}$. Nonadjacent picked momenta give marked edges with disjoint endpoints.

To construct regular points, give the endpoints of each marked edge opposite values of an independent parameter. Give each remaining vertex its own parameter. Denote the resulting values by $z_i$ and assign
\begin{equation}
 X_{ab}=z_a+z_b,\qquad z_{i+1}=-z_i\quad\text{if }p_i\text{ is picked}.
 \label{eq:boundary-witness}
\end{equation}
The four terms in every block equation~\eqref{eq:blockmesh} cancel. Any boundary corner is a separator edge, where $z_i+z_{i+1}=0$ already holds. Every physical channel is a nonzero linear form in two independent parameters, since no physical diagonal joins the endpoints of a marked edge. Generic parameter values therefore make all channels nonzero. These assignments establish regularity and are not additional conditions on the zero. They also establish regularity if extra complete-block rectangles between picked separators are imposed.

The same assignments restrict which poles can coincide after imposing the zero conditions. If two channel forms are proportional on the full locus, they must be proportional on Eq.~\eqref{eq:boundary-witness}. They must therefore involve the same two independent endpoint parameters. Replacing just one endpoint by its marked partner, or moving two endpoints in opposite directions, violates the physical length condition modulo $m-2\geq3$. The only possible distinct proportional channels are
\begin{equation}
 X_{ab},\qquad X_{a+1,b+1}=-X_{ab},
 \label{eq:density-opposite}
\end{equation}
with $p_a,p_b$ picked, or the same pair read in reverse. These two diagonals cross. No compatible tree contains both, so all restricted poles are simple. Moreover, a channel internal to either cut polygon cannot be proportional to its attaching channel. Thus the lower amplitudes have nonzero denominators at a generic point on the cut.

\subsection{What must survive a factorization cut}
Cut a physical diagonal $(a,b)$ and call the two sides $L$ and $R$, with $V_L+V_R=V$. Their original boundary paths contain $(m-2)V_L+1$ and $(m-2)V_R+1$ momenta. Each lower amplitude acquires one additional external momentum from the cut. On the left it is
\begin{equation}
 k_L=-\sum_{r=a}^{b-1}p_r=\sum_{r=b}^{a-1}p_r,
 \qquad k_R=-k_L,
 \label{eq:density-cut-momentum}
\end{equation}
where the second sum runs cyclically along the right path. On $X_{ab}=0$, both cut momenta are massless. Before imposing the zero conditions, ordinary tree factorization gives
\begin{equation}
 \mathop{\rm Res}_{X_{ab}=0}A_n^{(m)}=A_L^{(m)}A_R^{(m)}.
 \label{eq:density-residue}
\end{equation}
To make this residue vanish, it is enough to find an admissible picked set on one side and show that every block condition of its lower zero follows from the parent conditions.

Let $K_L,K_R$ count picked momenta on the two original paths, excluding the new cut momenta. They obey
\begin{equation}
 K_L+K_R=qV+2.
 \label{eq:density-pick-count}
\end{equation}
There are only two possibilities. One side already has the $qV_i+2$ picks required for a lower zero, or each side is short by exactly one. We treat them in turn.

\subsection{One side already has enough picks}
Suppose $K_L\geq qV_L+2$. Keep a consecutive group of exactly $qV_L+2$ picks in their order along the left path and leave $k_L$ unpicked. The group is consecutive among the picks, not among all momenta. This gives an admissible lower picked set. It requires $V_L\geq2$, since a one-vertex path has room for only $q+1$ nonadjacent picks.

Consider any retained separator pair of this lower zero. Its arc avoiding the cut contains no discarded pick, because the kept group was consecutive. The arc therefore has exactly the same length and picked count in the parent. The separator pair passes the parent density test as well, so its entire parent rectangle is imposed.

The only change in its block partition is at the block containing $k_L$. Replacing $k_L$ by the right path in Eq.~\eqref{eq:density-cut-momentum} replaces one momentum by $(m-2)V_R+1$ momenta. That lower block now contains
\begin{equation}
 (m-3)+\bigl[(m-2)V_R+1\bigr]=(m-2)(V_R+1)
\end{equation}
parent momenta. Since the replacement adds a multiple of $m-2$ entries, all other block boundaries remain aligned. The changed block is the sum of $V_R+1$ consecutive parent blocks. For any block $B$ on the other arc, its required contraction is consequently
\begin{equation}
 \widetilde B\cdot B
 =\left(\sum_{\nu=0}^{V_R}B^{(\nu)}\right)\!\cdot B
 =\sum_{\nu=0}^{V_R}B^{(\nu)}\cdot B=0.
 \label{eq:density-inherited-block}
\end{equation}
Every term is one of the imposed parent conditions. Blocks not containing the cut are unchanged. Thus every condition of the lower zero holds, and induction gives $A_L^{(m)}=0$. The argument is identical with $L$ and $R$ exchanged.

\subsection{Each side needs one more pick}
If neither side has enough picks, Eq.~\eqref{eq:density-pick-count} forces
\begin{equation}
 K_L=qV_L+1,\qquad K_R=qV_R+1.
 \label{eq:density-balanced-cut}
\end{equation}
We try to supply the missing pick by selecting the cut momentum. On $L$, this is allowed when neither $p_a$ nor $p_{b-1}$ is picked, since these are its two neighbours. On $R$, the corresponding neighbours are $p_b$ and $p_{a-1}$. Nonadjacency of the original picks implies that both choices fail only when $p_a,p_b$ are picked, or when $p_{a-1},p_{b-1}$ are picked. These are the paired poles treated in the next subsection.

Otherwise select the cut on an allowed side, say $L$, and retain all old picks there. This supplies exactly $qV_L+2$ picks. Again $V_L\geq2$, because a single $m$-gon cannot hold this many nonadjacent picks. A lower rectangle between two old separators is inherited exactly as in the preceding subsection. It remains to consider a rectangle whose separators are the new cut and an old picked momentum $p_c$.

\emph{Finding an original separator to replace the cut.} The right path contains $(m-2)V_R+1$ momenta and $qV_R+1$ picks. We need one of these picks to divide that path into two complete-block runs with the required density. The following elementary count guarantees it. On a path of $(m-2)W+1$ momenta with $K=qW+1$ picks, number the picked positions from zero as $t_1<\cdots<t_K$ and set
\begin{equation}
 H_j=(m-2)(j-1)-qt_j.
\end{equation}
The first value is $H_1=-qt_1\leq0$, and the last is $H_K=q[(m-2)W-t_K]\geq0$. Nonadjacency gives
\begin{equation}
 H_{j+1}-H_j=(m-2)-q(t_{j+1}-t_j)\leq1.
\end{equation}
An integer sequence starting nonpositive and ending nonnegative, with no upward step larger than one, must pass through zero. Since $q$ and $m-2=2q+1$ are coprime, a zero gives
\begin{equation}
 t_j=(m-2)t,\qquad j-1=qt,\qquad 0\leq t\leq W.
 \label{eq:density-path}
\end{equation}
Removing this picked momentum leaves runs of lengths $(m-2)t$ and $(m-2)(W-t)$, with $qt$ and $q(W-t)$ picks. An empty run at an endpoint is allowed.

Apply this fact with $W=V_R$ and call the resulting picked momentum $p_d$. The two arcs of the lower rectangle between $p_c$ and $k_L$ have lengths $(m-2)h$ and $(m-2)(V_L-h)$ and picked counts $qh$ and $q(V_L-h)$. Restoring the right path appends one of the runs ending at $p_d$ to each arc. With $t$ chosen for the first run, the parent arcs between $p_c,p_d$ therefore have lengths and picked counts
\begin{equation}
 \begin{array}{c|cc}
 &\text{length}&\text{picked momenta}\\\hline
 \text{first arc}&(m-2)(h+t)&q(h+t)\\
 \text{second arc}&(m-2)(V-h-t)&q(V-h-t).
 \end{array}
 \label{eq:density-extended-arcs}
\end{equation}
Thus $(c,d)$ is retained in the parent. Each appended run contains a whole number of blocks, so every block in the lower rectangle is an unchanged block in this parent rectangle. All its contractions vanish. This completes the inheritance argument also when the cut is picked.

Induction needs only that the inherited lower kinematics satisfy the lower zero conditions and have generically nonzero denominators. They need not cover the entire lower zero locus, because the lower amplitude vanishes at every regular point of that locus.

\subsection{Cancellation of the paired poles}
The only remaining case has the counts~\eqref{eq:density-balanced-cut} and picked momenta $p_a,p_b$, or equivalently the reverse shifted pair. Removing $p_a$ from the left path leaves $(m-2)V_L$ momenta and $qV_L$ picks. Thus $(a,b)$ passes the density test. Its rectangle relates the two crossing channels $X_{ab}$ and $X_{a+1,b+1}$.

To see that it relates their residues as well, label the block boundaries on the two arcs by
\begin{equation}
 u_r=a+1+r(m-2),\quad 0\leq r\leq V_L,
 \qquad v_s=b+1+s(m-2),\quad0\leq s\leq V_R.
\end{equation}
Labels are cyclic, so $u_{V_L}=b$ and $v_{V_R}=a$. Each tile sets one mixed difference of $X_{u_r,v_s}$ to zero. Summing the tile equations from any corner to the endpoints gives
\begin{equation}
 X_{u_r,v_s}=X_{a,u_r}+X_{b,v_s}-X_{ab}.
 \label{eq:density-paired-array}
\end{equation}
Here adjacent boundary channels vanish. Setting $r=s=0$, or just one of them to zero, yields
\begin{equation}
 \begin{gathered}
 X_{a+1,b+1}=-X_{ab},\\
 X_{b+1,u_r}=X_{a,u_r}-X_{ab},\qquad
 X_{a+1,v_s}=X_{b,v_s}-X_{ab}.
 \end{gathered}
 \label{eq:density-paired-channels}
\end{equation}
The interior values of $u_r$ and $v_s$ exhaust the allowed internal diagonals incident on $a$ and $b$ in their respective cut polygons. At $X_{ab}=0$, these equations identify their values with those of the shifted cut, while all other internal diagonals are unchanged. The two products of complete lower amplitudes are therefore equal. Their attaching propagators have opposite signs, so their residues cancel. The earlier classification of coincident poles ensures that no lower denominator vanishes identically in this comparison.

\subsection{Completion of the induction}
At $V=2$, every polygon vertex belongs to a marked edge, so the picks alternate. The retained rectangles pair the $m-1$ exchange channels with their opposite values, and the amplitude vanishes. This starts the induction.

For larger $V$, every pole is covered by one of the preceding cases. Either one lower amplitude vanishes by induction, or two equal factorization numerators cancel against opposite propagators. If a proportional partner also occurs in the first case, each residue already vanishes separately. Thus every finite pole of the restricted amplitude cancels.

The result is a polynomial in linear coordinates on the zero locus. However, every tree term has homogeneous degree $-(V-1)<0$ in those coordinates. Such a polynomial must be zero. Together with regularity, this proves the claim at every allowed multiplicity and every odd $m\geq5$. The $m=3$ restriction is the established cubic rectangle theorem. This argument proves vanishing of the physical amplitude, not uniqueness of a trial ansatz.

\subsection{An optional relaxation of width-one strips}
\label{sec:summed-strips}
A particular picked set admits larger zeros obtained by combining block conditions. Pick every other position along a consecutive stretch of the cyclic ordering,
\begin{equation}
 \{p_n,p_2,p_4,\ldots,p_{n-V}\}.
\end{equation}
Write $m-2=2q+1$, with $q=(m-3)/2$. For this picked set, the equal-density rule selects width-one rectangles with rows
\begin{equation}
 c=1+2s,\qquad s=0,\ldots,q(V-1),
\end{equation}
and columns $j=1,\ldots,V-1$. The entry in row $c$, column $j$ is $B_c\cdot B_{c+1+j(m-2)}$.

To specify the relaxation, place boundaries before column $1$ and after column $V-1$. Retain an interior boundary after column $j$, with $1\leq j\leq V-2$, precisely when
\begin{equation}
 \bigl[j=1\ \text{and}\ s\leq q(V-2)\bigr]
 \quad\text{or}\quad
 s+qj=q(V-1).
 \label{eq:summed-strip-cuts}
\end{equation}
Consecutive retained boundaries divide the columns into groups. If two consecutive boundary positions are $u<v$, set $I=\{u+1,\ldots,v\}$ and impose only
\begin{equation}
 \boxed{B_c\cdot\left(\sum_{j\in I}B_{c+1+j(m-2)}\right)=0.}
 \label{eq:summed-strip-groups}
\end{equation}
Impose this equation for every group in every row. Thus a group with one column gives an individual block condition, while a larger group allows cancellations among its contractions.

These grouped conditions give zeros at every allowed multiplicity for odd $m$. The grouping is equivalent to a full-row sum together with a prefix sum at each retained interior boundary. Summing Eq.~\eqref{eq:blockmesh} over a prefix telescopes to the corresponding channel relation. Regularity follows from the regular points of the stronger pairwise locus. At $m=3$, $q=0$, every interior boundary is retained. No blocks are combined, and the usual cubic width-one zero is recovered.

The cancellation uses the ordinary current recursion~\eqref{eq:BG}, with the row starts treated from largest to smallest. Opposite shortest channels remove child currents in pairs. Each omitted prefix condition multiplies a lower current already forced to vanish by the later rows, so it need not be imposed separately. Since $m-3$ is even, this pair removal terminates. Before restoring the full-row condition at $c=1$, the root sum reduces to
\begin{equation}
 A_n^{(m)}=\left(\frac1{X_{1,m}}+\frac1{X_{m-1,n}}\right)\mathcal A_{n-m+2}^{(m)},
 \label{eq:summed-root-cancellation}
\end{equation}
where $\mathcal A_{n-m+2}^{(m)}$ is the complete lower tree with $(p_n,p_1,\ldots,p_{m-2})$ merged into one generally off-shell leg and all internal channels inherited. The remaining full-row sum gives $X_{m-1,n}=-X_{1,m}$ and cancels the prefactor. This current argument is separate from the factorization proof of the general pairwise prescription above.

For $\tr(\phi^5)$ at fourteen points, $q=1$ and $V=4$. The four rows give
\begin{equation}
 \begin{aligned}
 B_1\cdot B_5&=B_1\cdot(B_8+B_{11})=0,\\
 B_3\cdot B_7&=B_3\cdot B_{10}=B_3\cdot B_{13}=0,\\
 B_5\cdot B_9&=B_5\cdot(B_{12}+B_1)=0,\\
 B_7\cdot(B_{11}+B_{14}+B_3)&=0.
 \end{aligned}
 \label{eq:summed-strip-fourteen}
\end{equation}
The cyclic labels have been reduced modulo fourteen. For example, $B_1\cdot B_8$ and $B_1\cdot B_{11}$ can be nonzero and opposite. This is an additional relaxation of the displayed picked set. The general equal-density prescription and the reconstruction checks in the main text retain every selected pairwise block condition.

\section{Rod zeros, regularity, and uniqueness with locality}
\label{sec:product}

\subsection{Why the packing rules give a zero}
\label{sec:product-current-proof}
The interval rule makes every proper current equal the weight of the complete packing on its interval. The root amplitude then becomes the sum over packings with one fewer rod, which the closing rule sets to zero. We establish these two statements to prove
\begin{equation}
 A_n^{(m)}\big|_{\G}=0,\qquad m\geq4,\quad V\geq2.
 \label{eq:product-zero-main}
\end{equation}
Let $J_{[a,b)}$ be the current including its attaching propagator. It obeys the ordinary tree recursion~\cite{BG},
\begin{equation}
 X_{ab}J_{[a,b)}=\sum_{[a,b)=I_1\sqcup\cdots\sqcup I_{m-1}}\prod_rJ_{I_r},
 \qquad J_{[a,a+1)}=1.
 \label{eq:BG}
\end{equation}
Each child has $1\bmod (m-2)$ leaves. For $m\geq4$, define trial currents by
\begin{equation}
 b_{[a,b)}=\prod_{r=0}^{\ell-1}X_{a+(m-2)r,a+(m-2)(r+1)+1}^{-1},
 \qquad b-a=(m-2)\ell+1,\qquad b_{[a,a+1)}=1.
 \label{eq:trial-main}
\end{equation}
The trial current $b_{[a,b)}$ is precisely the complete packing weight $S_\ell(a,b-m+1)$ in Eq.~\eqref{eq:closed-long-main}. For a single rod, the ratio in that equation is $1/(1/X)=X$. For several rods, division by the complete-packing weight cancels unchanged factors and gives the dimension of a channel.

A partition into $m-1$ children contains a total of $\ell-1$ vertices below its first vertex. Each product of trial currents corresponds to a unique packing of $\ell-1$ rods, and conversely. To see this, write the rod starts as
\begin{equation}
 i_r=a+(m-2)(r-1)+d_r,\qquad
 0\leq d_1\leq\cdots\leq d_{\ell-1}\leq m-2.
 \label{eq:offset}
\end{equation}
The offset $d_r$ counts the cuts preceding rod $r$ and therefore specifies its child interval. The number of rods with each offset $0,\ldots,m-2$ gives the number of vertices in that child. This explains the correspondence in both directions. The sum of child products is consequently $S_{\ell-1}(a,b-m+1)$. Dividing this sum by the complete-packing weight $b_{[a,b)}$ gives exactly the interval channel in Eq.~\eqref{eq:closed-long-main}. Thus Eq.~\eqref{eq:BG} proves $J_I=b_I$ by induction. At the amputated root, the same counting gives
\begin{equation}
 \left.A_n^{(m)}\right|_{\text{longer-channel relations}}=S_{V-1}(1,n-m+1).
 \label{eq:rod-root-proof}
\end{equation}
Separating these packings according to whether they contain the final rod gives
\begin{equation}
 \begin{aligned}
 S_{V-1}(1,n-m+1)
 &=S_{V-1}(1,n-m)+
 \frac{S_{V-2}(1,n-2m+3)}{X_{n-m+1,n}}\\
 &=0\qquad\text{on }\G.
 \end{aligned}
 \label{eq:rod-balance-main}
\end{equation}
The two terms omit or include that rod. Their cancellation follows from the assignment~\eqref{eq:closed-main}, proving that the amplitude vanishes. Singling out the last rod is only a parametrization choice.

\subsection{Regular points at every multiplicity}
To prove that the regular part is nonempty, it suffices to exhibit one point. For this purpose only, set every unfixed shortest channel to one in a common unit of momentum squared. Counting packings with and without the last rod gives $X_{n-m+1,n}=-(V-1)/(m-2)$. For any proper interval $b-a=(m-2)\ell+1$ the resulting channel is
\begin{equation}
 X_{ab}=\begin{cases}
 \displaystyle\binom{\ell+m-3}{m-2},&b<n,\\[4pt]
 \displaystyle-\frac{V-\ell}{m-2}\binom{\ell+m-4}{m-3},&b=n.
 \end{cases}
 \label{eq:regular}
\end{equation}
Both are nonzero for $1\leq\ell<V$. The second expression follows by separating placements that contain the last rod, while the first counts all placements on the interval. Thus every allowed $m,n$ has a regular rod zero, and the $n-m$ unfixed shortest channels vary in a nonempty open set.

\subsection{Root choices and the number of conditions}
\label{sec:rod-root-count}
At a fixed root, the $n-m$ unfixed shortest channels are coordinates on the regular rod locus. Each remaining physical channel is determined by one equation. Thus the $n(V-1)/2$ physical channels give the independent-condition count in Eq.~\eqref{eq:rod-condition-count}. A different choice of channel for solving the closing equation only changes the parametrization. Reversing the rooted line also gives the same packing rules.

For $m=4$, cyclic invariance of the rank-one condition makes all roots equivalent. At $V=2$, every root gives the same condition $\sum_eX_e^{-1}=0$. For $m\geq5$ and $V\geq3$, we can distinguish all $n$ roots using the regular point~\eqref{eq:regular} rooted at $n$. Define, with cyclic labels,
\begin{equation}
 D_a=X_{a,a+2m-3}
 -X_{a,a+m-1}X_{a+m-2,a+2m-3}
 \sum_{j=0}^{m-2}\frac1{X_{a+j,a+j+m-1}}.
\end{equation}
This is the first longer-channel rod relation. A rooted locus requires $D_a=0$ whenever its root is outside the consecutive run of $2m-3$ momenta starting at $a$. Substituting Eq.~\eqref{eq:regular} gives
\begin{equation}
 \begin{aligned}
 D_{n-2m+4}&=\binom{V+m-5}{m-2}-1
 -\binom{V+m-4}{m-2}\left(m-3-\frac{m-2}{V-1}\right)<0,\\
 D_{n-m+1}&=V-1-\binom{V+m-5}{m-3}<0,\\
 D_n&=\binom{V+m-5}{m-3}-1-\frac{2}{m-2}\binom{V+m-6}{m-3}>0.
 \end{aligned}
\end{equation}
The signs hold for all $m\geq5,V\geq3$. For the first, the ratio of its two binomial coefficients is $(V-2)/(V+m-4)<(V-2)/(V-1)\leq m-3-(m-2)/(V-1)$. For the second, the binomial coefficient is at least $V(V-1)/2$. The third expression increases with $V$ by Pascal's identity and is already $(m-4)(m-1)/(m-2)>0$ at $V=3$.

The three corresponding momentum runs intersect only at leg $n$. Every other root therefore requires at least one of these nonzero $D_a$ to vanish. The regular point belongs only to its original rooted locus, and cyclic relabelling proves that all $n$ root choices give distinct loci. This counts the cyclic rod construction, without asserting completeness among all amplitude zeros.

\subsection{Uniqueness when locality is assumed}
\label{sec:uniqueness}
Take the tree ansatz~\eqref{eq:localansatz} with constant coefficients. An end vertex is attached to the rest of the tree by one internal line and carries $m-1$ consecutive external legs. We first use limits within the rod locus to equate the coefficients of all diagrams containing a specified end vertex. Diagrams containing two such vertices then connect these groups of coefficients.

\emph{Starting the induction.} At $V=2$, the rod family covers $\sum_eX_e^{-1}=0$. In the variables $1/X_e$, this is a hyperplane. A linear combination $\sum_e c_e/X_e$ vanishes on it only when all $c_e$ agree.

For the induction, express the shortest channels as $X_{i,i+m-1}=w_i\cdots w_{i+m-3}$. This introduces no restriction on them. One can leave the first $m-3$ nonzero weights arbitrary and solve the consecutive products successively for the remaining weights. The rod assignments become
\begin{equation}
 X_{ab}=K_{ab}(w)=
 \sum_{0\leq r_1\leq\cdots\leq r_{m-2}<\ell}
 \prod_{s=1}^{m-2}w_{a+s-1+(m-2)r_s},
 \qquad b-a=(m-2)\ell+1,
 \qquad K_{1n}=0.
 \label{eq:K}
\end{equation}
Indeed, in the recursion for trial currents, multiplying by all gap weights cancels every factor except the $m-2$ cuts between children, yielding this sum.

\emph{An end vertex away from the root.} For a start $2\leq a\leq n-m$, scale the $m-2$ gap weights of its channel as $\ee\alpha_1,\ldots,\ee\alpha_{m-2}$, with every $\alpha_i\ne0$. Deleting these gaps at $\ee=0$ leaves arbitrary lower rod data. The closing equation can be maintained by adjusting a noninserted weight, since it is linear in that weight and has a nonzero derivative at generic regular lower data. Every longer channel has a nonzero limit given by a channel of the smaller polygon. The chosen end propagator has order $\ee^{-(m-2)}$. Compatible shortest propagators occupy disjoint weight segments separated by an unused gap, so diagrams omitting the chosen one cover at most $m-3$ scaled gaps. Thus
\begin{equation}
 0=\lim_{\ee\to0}\ee^{m-2}\widetilde A_n
 =\frac{\widetilde A_{n-m+2}^{(a)}|_{\G_{n-m+2}}}{\alpha_1\cdots\alpha_{m-2}}.
 \label{eq:earlimit}
\end{equation}
Here $\widetilde A_{n-m+2}^{(a)}$ is obtained by deleting the selected end vertex from every diagram that contains it, retaining its original coefficient. The remaining diagrams are precisely the lower compatible tree ansatz. The displayed limit says that this lower ansatz vanishes on the lower rod family. Induction therefore equates all its coefficients, without assuming their factorization values.

\emph{End vertices next to the root.} The boundary requires a separate limit because two crossing channels vanish together. Write the final weights as $(t_1,\ldots,t_{m-3},z)$ and $P=\prod_jt_j$. Writing $F_V=K_{1n}$ and $F_{V-1}$ for the root polynomial of the first $V-1$ blocks gives $F_V=F_{V-1}+zR$, where
\begin{equation}
 \frac RP=1+\sum_{k=1}^{m-3}\frac{Q_k}{t_1\cdots t_k},\qquad
 Q_k=\sum_{1\leq i_1\leq\cdots\leq i_k\leq V-1}
 \prod_{j=1}^kw_{(m-2)(i_j-1)+j}.
\end{equation}
Along $F_V=0$, the crossing channels $X_{n-m+1,n}=zP$ and $X_{1,n-m+2}=-zR$ vanish together. A compatible diagram contains at most one of them. At $z=0$, deleting either end vertex leaves the same lower rod data, built from the first $V-1$ blocks of weights. Thus neither inherited lower function depends on the $t_j$. Denoting them by $\widetilde A_{\rm tail}$ and $\widetilde A_{\rm root}$, the $1/z$ coefficient gives
\begin{equation}
 \widetilde A_{\rm root}=\frac{R}{P}\widetilde A_{\rm tail}.
\end{equation} For $m\geq4$, $R/P$ varies with the $t_j$ at generic lower data. For example, setting the weights with residues $1,\ldots,m-3$ modulo $m-2$ to one makes $Q_{m-3}$ a positive integer. Thus both lower functions must vanish separately. Induction therefore fixes both boundary groups, and reversing the ordering fixes the opposite boundary.

\emph{Connecting the groups.} A diagram with two compatible end vertices belongs to both groups, so their common coefficients must agree. For $V\geq4$, starts differing by at least $m-1$ connect every nonroot end vertex to one of the two extreme starts, which are themselves compatible. At $V=3$, the end vertex containing the root joins the otherwise isolated middle start to the others. Every tree has at least two end vertices and at most one contains the root. Hence all diagram coefficients agree. The proof requires $m\geq4$ and compatible simple poles. An overall normalization remains free.

\section{Real kinematics and threshold connections}
\label{sec:threshold-supp}
The classic threshold cancellation also has an interpretation in terms of reflectionless potentials~\cite{voloshinzeros,thresholdzeros}. The threshold configurations and the distinction between ordered and full amplitudes are discussed in Sec.~\ref{sec:physical-four}.

There is also a related connection in $\tr(\phi^3)$. The massive $2\to3$ threshold cancellation~\cite{makeenkothreshold} lies on the $\lambda=4$ member of the five-point zero family in Ref.~\cite{splitting}, expressed in inverse propagators, and admits a regular real continuation away from threshold. The following construction concerns the full $\tr(\phi^4)$ rank-one locus.

\subsection{Real momenta on the rank-one locus}
\label{sec:real-rank}
We now construct real momenta for any prescribed regular real rank-one channels. Write $\widehat X_{ab}=P_{ab}^{\,2}-M^2$ with common mass $M\geq0$. Factor the matrix as $C_{ij}=u_iv_j$. Its zero row and column sums allow periodic sequences satisfying $z_{i+1}-z_i=u_i$ and $s_{j+1}-s_j=v_j$. Integrating the channel mixed difference, with adjacent boundary channels set to zero, gives
\begin{equation}
 \widehat X_{2i+1,2j+2}=y_j-y_i-z_i(s_j-s_i),\qquad
 y_i-y_{i-1}=z_i(s_i-s_{i-1}),
 \label{eq:rank-secant}
\end{equation}
Here $y_i$ is also periodic, and the indices run cyclically from zero to $V$, where $n=2V+2$. For a forward interval, Eq.~\eqref{eq:rank-secant} equals $\sum_{i\leq r\leq k<j}u_rv_k$. Interleaving $w_{2r+1}=u_r$ and $w_{2r+2}=v_r$ identifies this with the $m=4$ rod formula~\eqref{eq:K}; the other endpoint parity gives the corresponding interchanged sum, and periodicity supplies the closing equation. Conversely, cyclic mixed differences of these rod sums give $C_{ij}=u_iv_j$. This explains the matrix and rod equivalence at all multiplicities.

The shortest channels are $\widehat X_{2i+1,2i+4}=(z_{i+1}-z_i)(s_{i+1}-s_i)$. Regularity therefore makes every consecutive difference of $s_i$ nonzero. Set
\begin{equation}
 z_i=\frac{y_i-y_{i-1}}{s_i-s_{i-1}},\qquad
 \rho>\max_j(y_j+s_j^2),\qquad r_j=\sqrt{\rho-y_j-s_j^2}.
\end{equation}
In signature $(+,-,-)$, define region momenta
\begin{equation}
 E_j=(0,s_j,r_j),\qquad O_i=(t_i,-z_i/2,0),\qquad
 t_i=\sqrt{(s_i+z_i/2)^2+r_i^2+M^2}.
\end{equation}
Direct subtraction gives
\begin{equation}
 (O_i-E_j)^2-M^2=y_j-y_i-z_i(s_j-s_i)
 =\widehat X_{2i+1,2j+2}.
 \label{eq:real-rank-identity}
\end{equation}
Set the external momenta to
\begin{equation}
 p_{2i+1}=O_i-E_i,\qquad p_{2i+2}=E_i-O_{i+1}.
\end{equation}
These momenta satisfy the three required checks. First, their sum telescopes to zero. Second, the adjacent channel values $\widehat X_{2i+1,2i+2}=\widehat X_{2i+3,2i+2}=0$ give $p_a^2=M^2$. Third, every physical propagator joins an odd region to an even region, so Eq.~\eqref{eq:real-rank-identity} reproduces every prescribed channel. All remain nonzero by the assumed regularity. Positive $t_i$ makes odd legs outgoing and even legs incoming. Adding a zero spatial component embeds the same configuration in $1+3$ dimensions.

\end{document}